\documentclass[aps,prl,twocolumn,groupedaddress,showpacs,floatfix]{revtex4}
\usepackage{graphicx}
\begin{document}
\title{The $^{40}$Ca($\alpha ,\gamma$)$^{44}$Ti
reaction in the energy regime of supernova nucleosynthesis}

\author {
H. Nassar$^1$,
M. Paul$^1$\footnote{{To whom correspondence should be addressed,
email address: paul@vms.huji.ac.il}},
I. Ahmad$^2$,
Y. Ben-Dov$^3$,
J. Caggiano$^2$,
S. Ghelberg$^1$,
S. Goriely$^4$,
J.P. Greene$^2$,
M. Hass$^5$,
A. Heger$^{6,7}$,
A. Heinz$^2$,
D.J. Henderson$^2$,
R.V.F. Janssens$^2$,
C.L. Jiang$^2$,
Y. Kashiv$^1$,
B.S. Nara Singh$^5$,
A. Ofan$^1$,
R.C. Pardo$^2$,
T. Pennington$^2$,
K.E. Rehm$^2$,
G. Savard$^2$,
R. Scott$^2$
and R. Vondrasek$^2$
}

\affiliation{$^1$ Racah Institute of Physics,
Hebrew University, Jerusalem, Israel, 91904}
\affiliation{$^2$ Argonne National Laboratory, Argonne, IL 60439, USA}
\affiliation{$^3$ Physics Div., NRC-Negev, POB 9001 Beer-Sheva, Israel 84190}
\affiliation{$^4$ Universit\'{e} Libre de Bruxelles, CP-226,
1050 Brussels, Belgium}
\affiliation{$^5$ Dept. of Particle Physics, Weizmann Institute,
Rehovot, Israel 76100}
\affiliation{$^6$ Los Alamos National Laboratory, Los Alamos, NM 87545, USA}
\affiliation{$^7$ University of California at Santa Cruz, Santa Cruz, CA 95064, USA}
\bibliographystyle{unsrt}
\date{\today}

\begin{abstract}
The $^{44}$Ti(t$_{1/2}$= 
59 y) nuclide, an important signature of
supernova nucleosynthesis, has recently been
observed as live
radioactivity by $\gamma$-ray astronomy from the Cas A
remnant. We investigate 
in the laboratory the major $^{44}$Ti production
reaction, $^{40}$Ca($\alpha ,\gamma$)$^{44}$Ti (E$_{\rm cm}\sim
0.6$--1.2 MeV/$u$), by direct 
off-line counting of $^{44}$Ti nuclei.
The yield, significantly higher than inferred from previous
experiments, is 
analyzed in terms of a 
statistical model using microscopic nuclear inputs. 
The associated stellar rate
has important astrophysical consequences, increasing
the calculated supernova $^{44}$Ti yield by a factor $\sim2$ over
previous estimates and bringing it closer to Cas A observations.
\end{abstract}


\maketitle

The radionuclide $^{44}$Ti(t$_{1/2}$= 59 y)
is considered an important signature
of explosive nucleosynthesis
in  core-collapse
supernovae (SN) \cite{arn:96}, where multiple $\alpha$ capture
is the path for SN nucleosynthesis
from $^{28}$Si to $^{56}$Ni(Fe).
$^{44}$Ti is mainly produced during 
an $\alpha$-rich freeze-out phase, the ratio $^{44}$Ti/$^{56}$Ni being
sensitive to the explosion conditions.
Stellar production of $^{44}$Ti determines the abundance
of stable $^{44}$Ca 
and contributes to
that of $^{48}$Ti (fed by $^{48}$Cr on the $\alpha$-chain).
Live  $^{44}$Ti has
been directly observed
from a point source identified as Cassiopeia A (Cas A)
by 
$\gamma$- and X-ray telescopes ({\it CGRO, RXTE, BeppoSAX})
and very recently by the
INTEGRAL mission (see \cite{die:05,vin:05}). Cas A is
believed to be the remnant of a 
core-collapse SN
whose progenitor mass was in the range
22--25 M$_{\odot}$ (M$_{\odot}$ denotes a solar mass)
\cite{vin:04}.
Using known values of the distance and age of the remnant,
half-life of $^{44}$Ti 
and the combined
$\gamma$ flux from all observations,
an initial $^{44}$Ti
yield of 160$\pm\,60\,\mu$M$_{\odot}$
is implied \cite{vin:05}.
This value is larger by a factor of 2--10
than $^{44}$Ti yields calculated
in current models
({\it e.g.} \cite{woo:95,thi:96})
and various explanations have been
proposed \cite{vin:04,nag:98,moc:99,the:98}.
$^{44}$Ti $\gamma$-ray emission 
from SN1987A
in the near Large Magellanic Cloud galaxy,
the closest known SN remnant in the last two
centuries,
is below detection limits.
But its present lightcurve
is believed to be powered by $^{44}$Ti radioactivity
and the inferred initial $^{44}$Ti
yield is estimated to be 100--200 $\mu$M$_{\odot}$
(see \cite{die:05}), similar to that of Cas A.
Using the $^{56}$Ni yield of SN1987A
directly measured by $\gamma$-ray astronomy,
the implied $^{44}$Ti/$^{56}$Ni ratio is larger
by a factor $\sim$3 than estimated by
stellar calculations \cite{die:05}.
No other source of
$^{44}$Ti activity 
has been confirmed so far,
despite a number of candidates and
the improved sensitivity of the INTEGRAL $\gamma$-ray
telescope \cite{die:05}.
Although many nuclear reactions play roles in determining the
SN yield of $^{44}$Ti
\cite{the:98,son:00}, the major production reaction is $^{40}$Ca($\alpha
,\gamma$)$^{44}$Ti and its importance
has been emphasized \cite{hof:99}.  Experimental information about
this reaction is incomplete and theoretical estimates are made less
reliable by the suppression of dipolar 
$T$\,=\,0\,$\rightarrow$\,0  transitions in
self-conjugate ($N$=$Z$) nuclei.  The reaction was studied 
in the 70's by $\alpha$ bombardment of 
Ca targets
and prompt-$\gamma$ spectroscopy 
down to E$_{\rm cm}\sim$ 3 MeV, 
yielding energies, spins
and strengths of isolated resonances \cite{coo:77,dix:77}.  Rauscher
{\it et al.} \cite{rau:00} 
used this information (together with
that for lighter $N$=$Z$ nuclei) to build an empirical model and
calculate an astrophysical rate for the reaction, currently adopted in
SN-nucleosynthesis codes.  The aim of the present work is to provide
new experimental information on the cross section of the
$^{40}$Ca($\alpha ,\gamma$)$^{44}$Ti reaction and on the astrophysical
rate of production of $^{44}$Ti in the energy range of SN
nucleosynthesis.  We use 
accelerator mass
spectrometry (AMS) with which we determine 
the integral number of
ground-state residual nuclei produced in an activation run.
This method
has been 
used to
measure cross sections of astrophysical interest producing long-lived
nuclides \cite{pau:80,nas:05s}.
We measure 
for the first time the yield of the
$^{40}$Ca($\alpha ,\gamma$)$^{44}$Ti reaction integrated over a
large range of energies in the explosive-nucleosynthesis regime,
allowing us 
to validate or scale the results of current
theoretical models and derive an astrophysical rate compatible
with laboratory experiment.

Our measurements were performed by bombarding a He-gas target cell with a
$^{40}$Ca beam and implanting forward-recoiling reaction products in a
Cu catcher acting also as beam stop. 
This inverse-kinematics approach has the advantage
of using a high-purity target able to sustain high beam power,
where the thickness
can be accurately
controlled by monitoring the gas pressure.
In the conditions of our experiment, the power deposited
in the gas ($<$10 mW/mm) is not expected to change
its stopping power \cite{goe:80}.
After irradiation, $^{44}$Ti atoms are chemically extracted
from the catcher with a $^{nat}$Ti carrier. The 
abundance $r_{44}= ^{44}$Ti/Ti (in the 10$^{-13}$-10$^{-12}$ range) is then
measured by AMS and the total number ($n_{44}$) of $^{44}$Ti nuclei produced
in the catcher is obtained by the relation $n_{44} = r_{44} n_{Ti}$, where
$n_{Ti}$ denotes the number of Ti carrier atoms used.
Importantly,
this result, based on the $^{44}$Ti/Ti isotopic ratio,
is independent of chemical
or counting efficiencies \cite{pau:80}.
The
experimental scheme was tested 
by producing a
copious number of $^{44}$Ti nuclei ($\sim 2.7\times 10^9$)
via the
reaction
$^{34}$S($^{12}$C,2n)$^{44}$Ti \cite{hui:00}. 
The $^{44}$Ti yield measured by AMS analysis and the
derived cross section  are consistent
with those calculated by the 
evaporation code PACE2 \cite{gav:80}.

The $^{40}$Ca irradiations were performed at two
different accelerators. Using the electron-cyclotron 
ion source
and the ATLAS linac at Argonne National Laboratory,
an intense
$^{40}$Ca$^{11+}$ beam ($\sim 1$ e$\mu$A on target) was produced.
The irradiation 
was 
set up in a 32 cm-diameter
chamber used as a gas cell.
The vacuum window, made of a 1.5 mg/cm$^2$ pinhole-free Ni foil
mounted eccentrically to the beam,
was rotated at $\sim$1 Hz
using a ferrofluidic vacuum seal
to dissipate
the power loss of the beam.
He gas
(99.995\%) was circulated 
at room temperature in the
large-volume chamber 
at constant
pressure (12 Torr)
during the $^{40}$Ca irradiation. 
The catcher,
at a distance of 18 cm from the entrance Ni foil,
was made of a 6.4 mm thick,
4.5 cm diameter
Cu (99.99 \%) disk,
clamped to a chilled-water cooled housing.
The beam intensity was periodically measured
during the irradiation run in an electron-suppressed
Faraday cup situated 2 m upstream of the gas cell 
and continuously
monitored by a beam scanner in front of the gas cell.
In addition, a small
Bragg-chamber detector
viewed the Ni window through a 0.8 mm diameter collimator
at a scattering
angle of 49$^{\circ}$ to
monitor ions elastically scattered off the foil.
Ion identification allowed for the discrimination between the main
$^{40}$Ca beam and a  (8.3$\pm$0.4)\%
contamination of
$^{40}$Ar.
A correction to the ion beam
charge was applied, using the monitored $^{40}$Ca fraction.
It should be noted that the $^{40}$Ar projectiles
cannot produce $^{44}$Ti from the $^4$He target.
The incident laboratory
energy (70.9 MeV)
and He pressure
(run R1, Table 1) were tuned
to populate a strong resonance (E$_{\rm cm}$= 4.10 MeV, J$^{\pi}= 2^+$)
in $^{44}$Ti, known from prompt $\gamma$-ray study \cite{dix:77}.
Using an Enge magnetic
spectrograph available on another beam line,
the energy of the $^{40}$Ca beam
after the Ni foil was 
measured before the irradiation (46.1 MeV, FWTM= 2.8 MeV).
The energy loss is well reproduced by calculations 
performed with the SRIM-2003 code \cite{zie:03}, which was 
used throughout our work to estimate 
specific energy loss.
\begin{figure}[t]
\includegraphics[width=60mm,height=60mm]{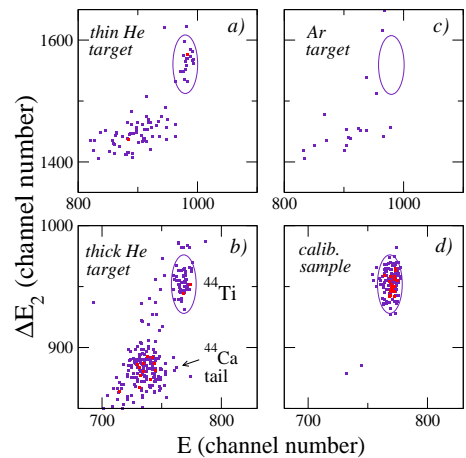}
\caption{\label{fig1} Identification spectra of $^{44}$Ti:
E and $\Delta$E$_2$ are energy and
energy-loss signals, respectively. The data shown
have been filtered through
a software condition set by additional energy-loss
signals ($\Delta$E$_1$, $\Delta$E$_4$) and correspond to
runs: {\it a)} R1; {\it b)} R2, R3; {\it c)}
B1; {\it d)} S1 (see Table 1).
}
\end{figure}
Irradiations under different conditions (see Table 1, Fig. 1)
were performed to verify that
production of $^{44}$Ti by background reactions is negligible:
in run B1,
the He gas was replaced by pure Ar (with similar incident energy loss)
and in run B2 (with He gas) the incident energy was
shifted downward ($\Delta$E$_{\rm cm}$= --\,0.25 MeV) off the resonance.
The other irradiations were performed at the
Koffler 
Tandem accelerator (Weizmann Institute), 
using a (isobarically pure)
$^{40}$Ca$^{8+}$ beam with an
average intensity of $\sim$120 enA on target.
The He (99.999\%) gas target and the water-cooled
Cu catcher were contained
in an electrically insulated and secondary-electron suppressed
chamber acting as a Faraday cup for beam charge integration.
Catchers made of the same high-purity Cu material and a
vacuum window made of the same Ni foil as in the ATLAS experiment,
were used.
In order to dissipate the power loss,
the beam  was magnetically steered 
onto the vacuum window in 
a periodic ($\sim$1\,Hz) spiral motion.
The incident energy of the beam
was set to 72.0\,MeV
and the thickness of the $^4$He gas target
(110\,Torr, 23\,cm) was
selected to integrate the reaction yield from E$_{\rm cm}$= 4.2 MeV
(after the vacuum window) down to  E$_{\rm cm}$= 2.1 MeV,
covering a large part of
the SN nucleosynthesis energy range.
Two independent irradiation runs were
performed under the same conditions, using different catchers
(R2, R3, see Table 1).

After activation, a 10 $\mu$m-thick layer was etched off
each Cu catcher in a 
HNO$_3$ solution containing 3 mg of Ti carrier.
Ti$^{4+}$ ions were separated from Cu
by ion-exchange techniques and Ti was finally 
converted to TiO$_2$.
The $^{44}$Ti analysis and measurement of the isotopic ratios
$r_{44}$ were performed using the Hebrew University AMS facility
\cite{ber:04} at the Koffler accelerator.
$^{44,46}$Ti$^-$ ions were alternately
accelerated at a terminal voltage
of 12 MV and 
$^A$Ti$^{10+}$ ions transported to the
detection system. $^{44}$Ti was discriminated from the
dominant $^{44}$Ca isobaric background
in a multi-anode ionization chamber
(Fig.~1) and the $^{46}$Ti$^{10+}$ charge current was
measured, 
leading to the 
$^{44}$Ti/Ti ratio.
In order to reduce systematic errors
in the final $r_{44}$ values, the ratios were
normalized to those measured for
a calibration sample 
(Table~1; 
see \cite{hui:00,ahm:98} for information on the 
$^{44}$Ti source material).
Blank samples (having gone through the same chemical procedures, but
containing no $^{44}$Ti) were used to
ensure that no background is present in the measurements.
More details on the AMS analysis
and the experiments can be found in \cite{ber:04,pau:03,
nas:05a}.
\begin{table}
\begin{center}
\caption{$^{40}$Ca irradiations and $^{44}$Ti AMS analysis results}
\begin{tabular}{lccccccr}
\hline
\hline
run &E$_{lab}$  & target/& irr.    & $^{40}$Ca & $^{44}$Ti  & $^{44}$Ti/Ti &  $n_{44}$ \\
    & (MeV)  &$\mu$g/cm$^2$ & acc.&   dose  &   AMS      &(10$^{-13}$)&  (10$^6$)  \\
      & ($^1$) &           & ($^2$) &(10$^{16}$)&  counts    & ($^3$)  &             \\
 \hline
 R1   & 46.1   & He/51  &   L     &    6.9(14)  &   20       &   1.7(5)     &  6.4(17) \\
 B1   & 46.1 & Ar/106 &   L     &    3.9(8)   &    0       &  $<0.1   $   &  $<0.3   $  \\
 B2   & 42.5 & He/42  &   L     &    2.6(5)   &    1       &  $0.1(1)$    &  $ 0.3(3)$  \\
 B3$^4$&  -  &   -    &   L     &      -      &    0       &  $<0.1   $   &      -      \\
 R2   & 46.4 & He/590&   T     &  3.06(12)    &   33       &   5.6(10)    & 21(4)   \\
 R3   & 46.4& He/590&   T     &  4.00(20)    &   28       &   7.7(15)    & 30(6)   \\
 S1$^5$&  -   &   -    &    -    &      -      &  377       &     380(20)     &     -    \\
 B4$^6$&   - &   -   &    -    &      -      &    0       &    $<0.1$     &       -     \\
\hline
\end {tabular}
\end{center}
\footnotesize {\raggedright
$^1$~$^{40}$Ca incident energy after pressure foil\\
$^2$~irradiation accelerator : L= linac, T= tandem  \\
$^3$~3.0 mg $^{nat}$Ti carrier used for each catcher \\
$^4$~unirradiated catcher \\
$^5$~calibration sample ($^{44}$Ti/Ti = (3.66$\pm$0.12)$\times$10$^{-11}$) \\
$^6$~blank AMS cathode ($^{nat}$Ti) \\
}
\end{table}

The experimental results are summarized in Table 1.
We derive from run R1
a resonance strength $\omega\gamma$
(see {\it e.g.} \cite{pau:03} for a relation between yield
and $\omega\gamma$) of
$8.8\pm3.0$ eV,
consistent
with the results of prompt-$\gamma$ measurements \cite{dix:77}
for the two close-by resonances in $^{44}$Ti at E$_x$= 9.227 and 9.239 MeV
($\omega\gamma\!=\! 5.8$ and 2.0 eV, respectively).
Runs R2 and R3
(thick He target)
give consistent $^{44}$Ti yields. If we interpret these yields as
the sum over isolated resonances in the experimental range of
energies, we derive lower and upper limits for the total resonance
strength of 30 eV and 63 eV, respectively, depending on the energies
of the resonances.
Experimentally known resonances \cite{dix:77,coo:77}
in this energy range,
and those used in the empirical model  \cite{rau:00},
sum to a total strength of 12.5 eV only.
Considerations based on the energy dependence of the Wigner limit 
for the $\alpha$ widths 
and of Weisskopf units for $\gamma$ transition intensities,
require that isolated resonances with total strength of the
order 30 to 60 eV ought to be centered
above E$_{\rm cm}$\,$\sim$\,3.6 MeV.
Such  resonances have not been observed
so far,
although this energy range has (at least in part) been investigated
by prompt-$\gamma$ spectroscopy. 
The additional yield measured in the present experiment may result
from unobserved resonances 
but more likely from closely spaced states in the $^{44}$Ti
compound nucleus (E$_x$($^{44}$Ti)=7.2--9.3 MeV),
unresolved in previous experiments.
We derive the experimental  average
cross section (thick He target) over the measured energy range
$\sigma^{exp}_{ave}= n_{44}/(N_{proj} n_T)\,=
\,(8.0\pm1.1)$ $\mu$b
($N_{proj}$ is the number of incident projectiles and $n_T$
the He target thickness in atoms/cm$^2$)
and compare it in
Fig. 2a with the expression
$\sigma_{ave}\!=\!
\int^{E_{max}}_{E_{min}} \sigma (E) (dE/dx)^{-1} dE/\Delta x$,
using values from current 
models for $\sigma (E)$.
The present data strongly support the BRUSLIB model \cite{arn:05}
which incorporates a
microscopic model of nuclear level densities \cite{dem:01}
and of the $\gamma$-ray strength function \cite{gor:02},
and a global $\alpha$-nucleus
optical-model potential \cite{dem:02}; 
the NON-SMOKER \cite{rau:01} and empirical \cite{rau:00} models
underestimate the average cross section.
In the BRUSLIB model, the $\gamma$ widths ($\Gamma_{\gamma}$)
are determined on the 
basis of a 
Quasi-Random Phase Approximation calculation (for details, 
see \cite{gor:02})
built on a Hartree-Fock description of the ground state, assuming
complete isospin mixing. 
For $N$=$Z$ nuclei, 
the standard prescription of ref.\,\cite{hol:76}
is adopted to account for the suppression of 
dipolar $T$\,=\,0\,$\rightarrow$\,0   transitions
by reducing the corresponding $\Gamma_\gamma$'s by a
factor $f_{iso}\!=\!5$.
The scaled BRUSLIB calculation (Fig. 2a) was modified to reproduce the
experimental average cross section by adjusting the suppression
factor to $f_{iso}\!=\!8$.
This value,
an experimental measure of the degree of mixing of the $T$\,=\,0
into the $T$\,=\,1 states, 
confirms the large reduction of $E1$
strength in a $N$\,=\,$Z$ nucleus.
It is close to the value ($f_{iso}\!=\!6-8$) derived 
in ref.\,\cite{har:86}
for the $^{24}$Mg and $^{32}$S compound nuclei
from a comparative analysis
of the cross sections of ($\alpha$,$\gamma$) reactions on 
$N$\,=\,$Z$   and ($Z$,$N$\,=\,$Z+2$) target nuclei.
The cross section $\sigma$(E) obtained from the scaled
BRUSLIB calculation was 
integrated to yield the astrophysical rate
of the $^{40}$Ca($\alpha , \gamma$)$^{44}$Ti reaction (Fig. 2b). The rate
is higher by a
factor 5-10 than that calculated with the model \cite{rau:00}
over the range of temperatures of
SN nucleosynthesis. It is well fitted by the expression
$N_A \langle\sigma v\rangle (cm^3 s^{-1} mole^{-1}) =
exp(107.92\!-\!2.254 x^{-1}\!+\!34.19 x^{-1/3}
        \!-\!167.81 x^{1/3}\!+\!6.867 x\!-\!0.2649 x^{5/3}
        \!+\!82.15 \ln(x))$
where $x=T_9$ in the range $0.1<\!T_9\!<10$
(T$_9$ denotes the temperature (K) divided by 10$^9$K). 
\begin{figure}[t]
\includegraphics[width=70mm,height=99mm]{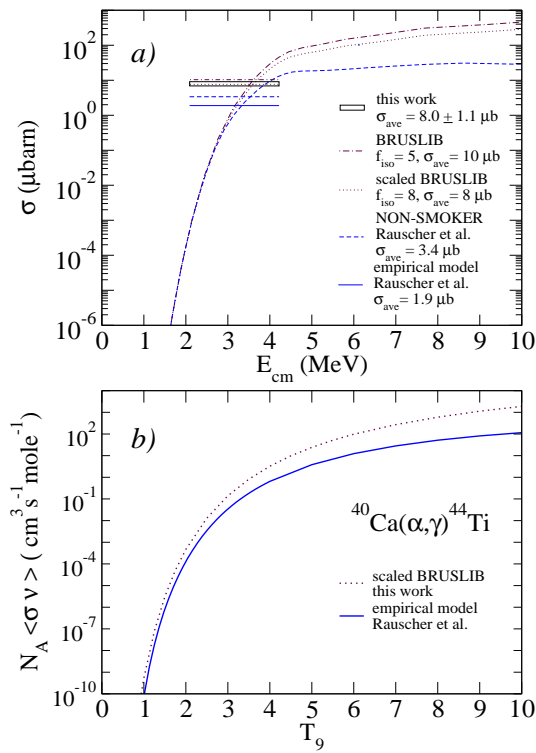}
\caption{\label{fig2} {\it a)} Comparison of the average cross section
of the $^{40}$Ca($\alpha , \gamma$)$^{44}$Ti reaction measured
in the present experiment (open box) with current models.
The horizontal bars represent the average cross section over the
energy range, calculated with the different models (see text);
{\it b)}
Astrophysical
rate of the reaction calculated with
scaled BRUSLIB (this work) and with the empirical model of Rauscher et al.
\cite{rau:00}, currently used in stellar calculations. See text.
}
\end{figure}

We have used this new rate 
in SN nucleosynthesis models and performed 
calculations as those
described in ref.\cite{rau:02}
with the KEPLER code \cite{wzw:78}.
For 15
and 25\,M$_{\odot}$  stars (Models S15 and S25), we find that,
while the $^{56}$Ni yield is kept at $0.1\,M_{\odot}$,
the SN
$^{44}$Ti yield in the ejecta increases
by a factor $\sim 2$ from $14\,\mu$M$_{\odot}$ to
$27\,\mu$M$_{\odot}$, and from 16$\,\mu$M$_{\odot}$ to
24$\,\mu$M$_{\odot}$, respectively.
Due solely to the new reaction rate, both the absolute 
$^{44}$Ti yield and the $^{44}$Ti/$^{56}$Ni ratio 
are thus brought closer to the values inferred
from $\gamma$-ray astronomy observations.
The higher estimated $^{44}$Ti/$^{56}$Ni ratio
is particularly important since it fits better
the solar $^{44}$Ca/$^{56}$Fe ratio.
Our calculations follow self-consistently 
spherical-symmetry hydrodynamics to determine 
the mass cut (the surface dividing the ejected mass from
that falling back onto the compact object). Increase
by a factor $\sim$2
in $^{44}$Ti is found also in the calculated fall back material.
The yields of multi-dimensional SN explosion calculations 
proposed to explain the observed $^{44}$Ti yield of
Cas A \cite{nag:98},
in which parts of deeper layers can be ejected while some of
the outer layers fall back, are thus expected to be 
enhanced in $^{44}$Ti as well.
The $^{48}$Cr
production increased 
in our calculations only by 13\,\% and 5.3\,\%,
increasing, after decay,  the total stellar 
$^{48}$Ti yield of Models S15 and S25 by
10\,\% and 4\,\%.


In summary, we have measured the integral yield and average cross
section of the $^{40}$Ca($\alpha,\gamma$)$^{44}$Ti reaction over a
large range of incident energies compatible with explosive
nucleosynthesis by 
off-line counting of $^{44}$Ti ground-state
nuclei. The yield
is well reproduced by statistical calculations
using microscopic nuclear inputs.  It would be
interesting to study the reaction in finer energy steps and
investigate whether some prominent resonances also
contribute to the yield.  The
associated astrophysical rate of the reaction
is significantly larger than the one adopted in
previous stellar calculations and increases the predicted SN $^{44}$Ti
yield by a factor of $\sim$2.
The new estimates may therefore account for the measured 
yield for the Cas A remnant and support the prospect of
identifying additional $^{44}$Ti localized sources in the Galaxy.


We gratefully acknowledge the participation of  
J.\ G\"orres, S.K. Hui and M. Wiescher in early stages of
this experiment.  We thank J.P. Schiffer, J. Truran and 
S. Woosley for stimulating discussions and
C.\ Feldstein and N.\ Trubnikov for their
work on the chemistry procedure.
This work was supported
in part by the US-DOE, Office of Nuclear Physics, under Contract No.
W-31-109-ENG-38, the DOE Program for Scientific Discovery through
Advanced Computing (SciDAC; DE-FC02-01ER41176), by DOE contract
W-7405-ENG-36 to the Los Alamos National Laboratory, and by the
USA-Israel Binational Science Foundation (BSF).


\end{document}